\documentclass[11pt,twoside,onecolumn]{article}
\usepackage[]{latexsym}
\usepackage{epsfig,upgreek}
\usepackage{amsmath,amssymb}
\RequirePackage[dvipsnames,usenames]{color}
\usepackage[colorlinks,hyperindex]{hyperref}

\definecolor{green1}{RGB}{0,128,0} 
 
\hypersetup{hidelinks,backref=true,pagebackref=true,hyperindex=true,colorlinks=true,breaklinks=true,urlcolor= blue}
\hypersetup{%
  colorlinks = true,
  linkcolor  = blue,
  citecolor = blue,
}
\usepackage{cite}
\newcommand{\be}{\begin{eqnarray}}
\newcommand{\ee}{\end{eqnarray}}

\renewcommand{\d}{\mbox{{\rm d}}}

\newcommand{\rH}{r_{\rm H}}
\newcommand{\rh}{r_{\rm h}}

\title{\bf Einstein-Klein-Gordon by gravitational decoupling}
\author{J.~Ovalle$^{ab}$\thanks{Corresponding author: jovalle@usb.ve}
$\,$, 
R.~Casadio$^{cd}$\thanks{casadio@bo.infn.it}
$\,$,
R.~da~Rocha$^{e}$\thanks{roldao.rocha@ufabc.edu.br} 
$\,$,
A.~Sotomayor$^{f}$\thanks{adrian.sotomayor@uantof.cl}
$\,$,
Z.~Stuchlik$^{a}$\thanks{zdenek.stuchlik@fpf.slu.cz}
\\
\null
\\
$^a${\em Institute of Physics and Research Centre of Theoretical Physics and Astrophysics,}
\\
{\em Faculty of Philosophy and Science, Silesian University in Opava}
\\
{\em CZ-746 01 Opava, Czech Republic}
\\
$^b${\em Departamento de F\'{\i}sica, Universidad Sim\'on Bol\'ivar,}
\\
{\em AP 89000, Caracas 1080A, Venezuela}
\\
$^c${\em Dipartimento di Fisica e Astronomia, Alma Mater Universit\`a di Bologna}
\\
{\em via Irnerio~46, 40126 Bologna, Italy}
\\
$^d${\em Istituto Nazionale di Fisica Nucleare, Sezione di Bologna, I.S.~FLAG}
\\
{\em  viale Berti~Pichat~6/2, 40127 Bologna, Italy}
\\
$^e${\em Centro de Matem\'atica, Computa\c c\~ao e Cogni\c c\~ao,}
\\
{\em Universidade Federal do ABC (UFABC)}
\\
{\em  09210-580, Santo Andr\'e, SP, Brazil.}
\\
$^f${\em Departamento de Matem\'aticas, Universidad de Antofagasta}
\\
{\em  Antofagasta, Chile}
}
\begin{document}
\maketitle
\begin{abstract}

We investigate how a spherically symmetric scalar field can modify the Schwarzschild vacuum solution when there is no
exchange of energy-momentum between the scalar field and the central source of the Schwarzschild metric.
This system is described by means of the gravitational decoupling by Minimal Geometric Deformation (MGD-decoupling),
which allows us to show that, under the MGD paradigm, the Schwarzschild solution is modified in such a way that a naked
singularity appears.
\end{abstract}
%
%\keywords{General relativity, Black holes}
%\pacs{04.50.+h, 04.70.-s, 04.70.Dy}
%
%
%
%
%
%
%\newpage
%
\section{Introduction}
\setcounter{equation}{0}
The Einstein-Klein-Gordon system has been studied for a long time.
In recent years, the interest in it has been renewed mainly because many alternatives to General Relativity (GR) contain
scalar fields (see Ref.~\cite{ts} for a brief review).
These scalar fields therefore appear of vital importance to elucidate the possible deviations suffered by GR.
If we want to study these deviations, the simplest system to be investigated is the spherically symmetric vacuum,
now converted into a scalar-vacuum, which could modify the well-known Schwarzschild black hole solution.
In this respect, possible conditions for circumventing the {\em no-hair\/} conjecture have been investigated
in different scenarios (see Refs.~\cite{Hawking:2016msc,thomas1,babi,thomas2,radu,marcelobh,thomas3,kanti1,adolfo2,adolfo3}
for some recent results and Refs.~\cite{galtsov1,kanti2,kanti3,galtsov2,mtz,konstantin} for earlier works).
In particular, a fundamental scalar field $\psi$ has been considered with great interest in Ref.~\cite{thomasBH}
(see also references therein).
\par
In a recent work~\cite{ocrss}, instead of considering specific fundamental fields to generate hair in black hole solutions,
we have assumed the presence of an additional completely generic source described by a {\em conserved\/}
energy-momentum tensor $\theta_{\mu\nu}$.
The main feature of the source $\theta_{\mu\nu}$, which defines a spherically symmetric ``tensor-vacuum'',
is that it gravitates but does not interact directly with the matter that sources the (hairless) black hole
solutions~{\footnote{This feature can be fully justified in the context of dark matter.}}.
This was developed in the context of the MGD-decoupling, originally proposed~\cite{jo1,jo2} in the context of the
brane-world~\cite{lisa2} and later extended to investigate new black hole solutions in
Refs.~\cite{MGDextended1,MGDextended2} (for some earlier works on the MGD, see for instance Refs.~\cite{jo6,jo8,jo9,jo10},
and Refs.~\cite{jo11,jo12,roldaoGL,rrplb,rr-glueball,rr-acustic,camiloleon,GLalysis,mgdanis,Luciano,ernesto1,sharif1,sharif2,angel2,sharif3,ernesto3,milko1,tello1,MGDBH,
emgd,MGDBH2,ernesto2}
for some recent applications).
\par
For a tensor-vacuum described above, many black black hole solutions with primary hairs and horizon
$r_{\rm H}=2\,M$ were found~\cite{ocrss}, and a fundamental characteristic was discovered:
the tensor-vacuum must be anisotropic.
This feature points to scenarios with Klein-Gordon (KG) type fields $\psi$, which naturally induce anisotropy in the Einstein
field equations.
This is precisely the case under scrutiny in this paper, namely, the tensor-vacuum will be represented
by the energy-momentum tensor $\theta_{\mu\nu}$ of a KG scalar field $\psi$, which gravitates
but does not interact directly with the matter that sources the central source of the Schwarzschild solution.
%\par
%{\color{red} EPL FORMAT DOES NOT SUPPORT NUMBERED SECTIONS, SO WE CANNOT REFER YO THEM:
%I WOULD JUST REMOVE THE FOLLOWING PARAGRAPH IN BLUE}
%{\color{blue}The paper is organised as follows:
%in Section~\ref{s2}, we present in details the MGD-decoupling applied to a spherically symmetric system containing
%two generic sources, namely, an energy-momentum tensor $T_{\mu\nu}$ and an additional one $\theta_{\mu\nu}$;
%in Section~\ref{s3}, we review briefly possible black hole solutions associated with the tensor-vacuum defined
%by $\{T_{\mu\nu}=0,\,\theta_{\mu\nu}\neq\,0\}$; the Einstein-KG system is then analysed in Section~\ref{s4}
%and bounds from the solar system tests are discussed in Section~\ref{s5}; 
%finally, we summarise our conclusions in Section~\ref{s6}.}
%
%
\section{MGD decoupling for two sources}
\label{s2}
\setcounter{equation}{0}
%
% 
%\textcolor{red}{
The MGD-decoupling represents the first simple and systematic approach to decoupling gravitational sources
in GR.
It has some properties very useful in the search for new solutions of Einstein's field equations.
The two main feature of this approach are the following~\cite{MGD-decoupling}:
\begin{itemize}
\item 
{\it Extending simple solutions into more complex domains.} We can start from a simple
energy-momentum tensor $T_{\mu\nu}$ and add to it more complex gravitational sources.
The starting source $T_{\mu\nu}$ could be as simple as we wish, including the vacuum indeed,
to which we can add a first new source, say
\be
\label{coupling}
T_{\mu\nu}\mapsto \tilde T^{(1)}_{\mu\nu}=T_{\mu\nu}+T^{(1)}_{\mu\nu}
\ .
\ee
We can then repeat the process with more sources.
In this way, we can extend solutions of the Einstein equations associated with the simplest gravitational
source $T_{\mu\nu}$ into the domain of more complex forms of gravitational sources
$T_{\mu\nu}=\tilde T^{(n)}_{\mu\nu}$, step by step and systematically.
This is precisely the case used in this paper: from the Schwarzschild solution ($T_{\mu\nu}=0$)
to the scalar-vacuum solution [see further Eq.~\eqref{KGem2}].
\par
\item {\it Deconstructing a complex gravitational source.}
In order to find a solution to Einstein's equations with a complex energy-momentum
tensor $\tilde T_{\mu\nu}$, we can split it into simpler components, say $T_{\mu\nu}$ and $T^{(i)}_{\mu\nu}$,
and solve Einstein's equations for each one of these parts.
Hence, we will find as many solutions as are the contributions $T^{(i)}_{\mu\nu}$ in $\tilde T_{\mu\nu}$.
Finally, by a straightforward combination of all these solutions, we will obtain the solution to the
Einstein equations associated with the original energy-momentum tensor $\tilde T_{\mu\nu}$.
\end{itemize}
Since Einstein's field equations are non-linear, the MGD-decoupling represents a powerful tool 
in the search and analysis of solutions, especially when we deal with situations beyond trivial cases.
\par
Let us start from the standard Einstein field equations
\begin{equation}
\label{corr2}
R_{\mu\nu}-\frac{1}{2}\,R\, g_{\mu\nu}
=
-k^2\,T^{\rm (tot)}_{\mu\nu}
\ ,
\end{equation}
and assume the total energy-momentum tensor has the form~\cite{Matt}
\begin{equation}
\label{emt}
T^{\rm (tot)}_{\mu\nu}
=
T_{\mu\nu}+\,\theta_{\mu\nu}
\ .
\end{equation}
Since the Einstein tensor satisfies the Bianchi identity, the total source in Eq.~(\ref{emt})
must satisfy the conservation equation
\begin{equation}
\nabla_\mu\,T^{{\rm (tot)}{\mu\nu}}=0
\ .
\label{dT0}
\end{equation}
\par 
Next we consider the static and spherical symmetric case, for which the metric $g_{\mu\nu}$ reads 
\begin{equation}
ds^{2}
=
e^{\nu (r)}\,dt^{2}-e^{\lambda (r)}\,dr^{2}
-r^{2}\,\d\Omega^2
\ ,
\label{metric}
\end{equation}
where $\nu =\nu (r)$ and $\lambda =\lambda (r)$ are functions of the areal
radius $r$ only, ranging from $r=0$ (the star center) to some $r=R$ (the
star surface). The metric~(\ref{metric}) must satisfy the Einstein equations~(\ref{corr2}),
which in terms of the two sources in~\eqref{emt} explicitly read,
\begin{eqnarray}
\label{ec1}
k^2
\left(
T_0^{\ 0}+\,\theta_0^{\ 0}
\right)
\!\!&\!\!=\!\!&\!\!
\frac 1{r^2}
-
e^{-\lambda }\left( \frac1{r^2}-\frac{\lambda'}r\right)\ ,
\\
\label{ec2}
k^2
\left(T_1^{\ 1}+\,\theta_1^{\ 1}\right)
\!\!&\!\!=\!\!&\!\!
\frac 1{r^2}
-
e^{-\lambda }\left( \frac 1{r^2}+\frac{\nu'}r\right)\ ,
\\
\label{ec3}
k^2
\left(T_2^{\ 2}+\,\theta_2^{\ 2}\right)
\!\!&\!\!=\!\!&\!\!
-\frac {e^{-\lambda }}{4}
\left(2\nu''+\nu'^2-\lambda'\nu'
+2\,\frac{\nu'-\lambda'}r\right)
\ \ \
\end{eqnarray}
where $f'\equiv \partial_r f$ and $\tilde{T}_3^{{\ 3}}=\tilde{T}_2^{\ 2}$ due to the spherical symmetry.
The conservation equation~(\ref{dT0}) is a linear combination of Eqs.~(\ref{ec1})-(\ref{ec3}), and yields
\begin{equation}
\label{con1}
\left(\tilde{T}_1^{\ 1}\right)'
-
\frac{\nu'}{2}\left(\tilde{T}_0^{\ 0}-\tilde{T}_1^{\ 1}\right)
-
\frac{2}{r}\left(\tilde{T}_2^{\ 2}-\tilde{T}_1^{\ 1}\right)
=
0
\ ,
\end{equation}
which in terms of the two sources in Eq.~\eqref{emt} read,
\begin{eqnarray}
\label{con11}
&
\left({T}_1^{\ 1}\right)'
-
\frac{\nu'}{2}\left({T}_0^{\ 0}-{T}_1^{\ 1}\right)
-
\frac{2}{r}\left({T}_2^{\ 2}-{T}_1^{\ 1}\right)
&
\nonumber
\\
&
+
\left[
\left({\theta}_1^{\ 1}\right)'
-
\frac{\nu'}{2}\alpha\left({\theta}_0^{\ 0}-{\theta}_1^{\ 1}\right)
-
\frac{2\,\alpha}{r}\left({\theta}_2^{\ 2}-{\theta}_1^{\ 1}\right)
\right]
=
0
\ .
&
\
\end{eqnarray}
\par
By simple inspection, we can identify in Eqs.~\eqref{ec1}-\eqref{ec3} an effective density $\tilde{\rho}$, an effective radial pressure $\tilde{p}_{r}$ and an effective tangential pressure $\tilde{p}_{t}$
\begin{equation}
\tilde{\rho}
=
T_0^{\ 0}
+\,\theta_0^{\ 0}\ ;\,\,\,\tilde{p}_{r}
=
-T_1^{\ 1}-\,\theta_1^{\ 1}
\ ;\,\,\,\tilde{p}_{t}
=
-T_2^{\ 2}-\,\theta_2^{\ 2}
\ ,
\end{equation}
which clearly show the anisotropy
\begin{equation}
\label{anisotropy}
\Pi
\equiv
\tilde{p}_{t}-\tilde{p}_{r}
\ .
\end{equation}
The system of Eqs.~(\ref{ec1})-(\ref{ec3}) may therefore be formally
treated as an anisotropic fluid~\cite{Luis,tiberiu,daniele,sante}.
\par
The MGD-decoupling can now be applied.
We can then proceed by considering a solution to the Eqs.~\eqref{corr2} for the source $T_{\mu\nu}$
[that is Eqs.~(\ref{ec1})-(\ref{con1}) with $\theta_{\mu\nu}=0$], which we can write as
\begin{equation}
ds^{2}
=
e^{\xi (r)}\,dt^{2}
-
\frac{dr^{2}}{\mu(r)}
-
r^{2}\left( d\theta^{2}+\sin ^{2}\theta \,d\phi ^{2}\right)
\ ,
\label{pfmetric}
\end{equation}
where 
\begin{equation}
\label{standardGR}
\mu(r)
\equiv
1-\frac{k^2}{r}\int_0^r x^2\,T_0^{\, 0}(x)\, dx
=
1-\frac{2\,m(r)}{r}
\end{equation}
is the standard GR expression containing the Misner-Sharp mass function $m(r)$.
The effects of the source $\theta_{\mu\nu}$ on $T_{\mu\nu}$ can then
be encoded in the MGD undergone solely by the radial component of the perfect fluid geometry
in Eq.~(\ref{pfmetric}).
Namely, the general solution is given by Eq.~\eqref{metric} with $\nu(r)=\xi(r)$ and
\begin{eqnarray}
\label{expectg}
e^{-\lambda(r)}
=
\mu(r)+\alpha\,f^{*}(r)
\ ,
\end{eqnarray} 
where $f^{*}$ is the (minimal) geometric deformation due to the effects of the source $\theta_{\mu\nu}$
and $\alpha$ a constant that helps to keep track of these effects.
Now let us plug the expression in Eqs.~(\ref{expectg}) in the Einstein equations~(\ref{ec1})-(\ref{ec3}).
The system is thus separated in two sets: i) one with the standard Einstein field equations for an
energy-momentum tensor $T_{\mu\nu}$, whose metric is given by Eq.~(\ref{pfmetric}), 
\begin{eqnarray}
\label{ec1pf}
k^2\,T_0^{\, 0}
\!\!&\!\!=\!\!&\!\!
\frac{1}{r^2}-\frac{\mu}{r^2}-\frac{\mu'}{r}\ ,
\\
\label{ec2pf}
k^2
\,T_1^{\, 1}
\!\!&\!\!=\!\!&\!\!
\frac 1{r^2}-\mu\left( \frac 1{r^2}+\frac{\xi'}r\right)\ ,
\\
\label{ec3pf}
k^2
\strut\displaystyle
\,T_2^{\, 2}
\!\!&\!\!=\!\!&\!\!
-\frac{\mu}{4}\left(2\xi''+\xi'^2+\frac{2\xi'}{r}\right)-\frac{\mu'}{4}\left(\xi'+\frac{2}{r}\right)
\end{eqnarray}
and its respective conservation equation
\begin{equation}
\label{conset1}
\left({T}_1^{\ 1}\right)'
-
\frac{\nu'}{2}\left({T}_0^{\ 0}-{T}_1^{\ 1}\right)
-
\frac{2}{r}\left({T}_2^{\ 2}-{T}_1^{\ 1}\right)
=
0
\ ,
\end{equation}
and ii) one with the equation of motion for the source $\theta_{\mu\nu}$, the so-called ``quasi-Einstein" system, which reads
\begin{eqnarray}
\label{ec1d}
k^2\,\theta_0^{\ 0}
\!\!&\!\!=\!\!&\!\!
-\frac{\alpha\,f^{*}}{r^2}
-\frac{\alpha\,f^{*'}}{r}\ ,
\\
\label{ec2d}
k^2\,\theta_1^{\ 1}
\!\!&\!\!=\!\!&\!\!
-\alpha\,f^{*}\left(\frac{1}{r^2}+\frac{\nu'}{r}\right)\ ,
\\
\label{ec3d}
k^2\,\theta_2^{\ 2}
\!\!&\!\!=\!\!&\!\!
-\frac{\alpha}{4}
\left[f^{*}\left(2\,\nu''+\nu'^2+2\frac{\nu'}{r}\right)
+f^{*'}\left(\nu'+\frac{2}{r}\right)\right]
\ \ \ \
\end{eqnarray}
and its conservation equation
\begin{equation}
\label{conset2}
\left({\theta}_1^{\ 1}\right)'
-
\frac{\nu'}{2}\left({\theta}_0^{\ 0}-{\theta}_1^{\ 1}\right)
-
\frac{2}{r}\left({\theta}_2^{\ 2}-{\theta}_1^{\ 1}\right)
=
0
\ .
\end{equation}
From the expressions~\eqref{con11},~\eqref{conset1} and~\eqref{conset2} we see that there is no
exchange of energy between the sources $T_{\mu\nu}$ and $\theta_{\mu\nu}$ and therefore
their interaction is purely gravitational.
\par
In the next section, we shall solve the above equations for a physically relevant and well-known system,
namely, the scalar-vacuum, described by the Einstein-KG equations.
\section{Black holes and naked singularities}
\label{s3}
Let us start by considering possible black hole solutions associated with the ``tensor-vacuum''
introduced in Ref.~\cite{MGDBH}, namely, a region of space containing a source described
by the energy-momentum $\theta_{\mu\nu}$ surrounding a self-gravitating system of radius $R$
and source $T_{\mu\nu}$.
The spherically symmetric region $r>R$ is thus described by Eqs.~\eqref{ec1d}-\eqref{ec3d} with
$T_{\mu\nu}=0$ and $\theta_{\mu\nu}\neq 0$.
In particular, the outer Schwarzschild metric generated by the compact source $T_{\mu\nu}$,
\begin{equation}
\label{Schw0}
ds^2
=
\left(1-\frac{2\,M}{r}\right)dt^2
-\left(1-\frac{2\,M}{r}\right)^{-1}dr^2
-r^2\,d\Omega^2\ ,
\end{equation}
will be modified by the the source $\theta_{\mu\nu}$ according to the expression~\eqref{expectg}, 
which now reads
\begin{eqnarray}
ds^2
\!\!&\!\!=\!\!&\!\!
\left(1-\frac{2\,M}{r}\right)dt^2
-\left[1-\frac{2\,M}{r}+\alpha\,f^{*}(r)\right]^{-1}
dr^2
\nonumber
\\
&&
-r^2\,d\Omega^2
\ ,
\label{Schw}
\end{eqnarray}
where the MGD function $f^{*}$ can be determined by imposing specific equations of state~\cite{MGDBH}
on the energy-momentum $\theta_{\mu\nu}$ to close the system of Eqs.~(\ref{ec1d})-(\ref{ec3d}).
\par
Let us recall that for the Schwarzschild metric~\eqref{Schw0} the surface $\rH=2\,M$
is both a Killing horizon, determined by $g_{tt}=e^\nu=0$, and an outer marginally trapped surface,
namely, the causal horizon where $g^{rr}=-e^{-\lambda}=0$.
For the MGD Schwarzschild metric~\eqref{Schw}, the hypersurface $r=\rH=2\,M$ is still a Killing horizon,
but contrary to the true vacuum $T_{\mu\nu}=\theta_{\mu\nu}=0$, it could become a real singularity
due to the effects of the energy-momentum $\theta_{\mu\nu}$.
On the other hand, the causal horizon is now located at $r=\rh$ such that 
\begin{equation}
\label{newBH}
\rh
\left[1+\alpha\,f^{*}(\rh)\right]
=
2\,M
\ .
\end{equation}
We should therefore require that $\rh\ge 2\,M$, so that the surface $r=\rH$ is hidden behind
(or coincides with) the causal horizon.
\par
In order to solve the equations of motion~\eqref{ec1d}-\eqref{ec3d} for the tensor-vacuum,
we will consider a generic equation of state~\cite{MGDBH} in the form
\begin{equation}
\label{generic}
\theta_0^{\,0}
=
a\,\theta_1^{\,1}+b\,\theta_2^{\,2}
\ ,
\end{equation}
with $a$ and $b$ constants. Eqs.~(\ref{ec1d})-(\ref{ec3d}) then yield the differential equation for the MGD function
\begin{eqnarray}
\label{giso}
&&\left(f^{*}\right)'
\left[\frac{1}{r}-\frac{b}{4}\left(\xi'+\frac{2}{r}\right)\right]
+f^{*}
\left[\frac{1}{r^2}-a\left(\frac{1}{r^2}+\frac{\xi'}{r}\right)\right.
\nonumber\\
&&\left.
-\frac{b}{4}\left(2\,\xi''+\xi'^2+2\frac{\xi'}{r}\right)\right]
=
0
\ ,
\end{eqnarray}
whose general solution for $r>\rH=2\,M$ is given by
\begin{equation}
f^*(r)
=
\left(1-\frac{2\,M}{r}\right)
\left(\frac{\ell}{r-B\,M}\right)^{A}
\ ,
\label{giso2}
\end{equation}
where $\ell$ is a positive constant with dimensions of a length, and
\begin{eqnarray}
\label{A}
A
\!\!&\!\!=\!\!&\!\!
\frac{2\,(a-1)}{b-2}>0
\\
\label{B}
B
\!\!&\!\!=\!\!&\!\!
\frac{b-4}{b-2}
\ ,
\end{eqnarray}
with $b\neq 2$ and the condition $A>0$ required by asymptotic flatness.
Therefore the MGD solution~\eqref{Schw} for the tensor-vacuum $\{T_{\mu\nu}=0,\,\theta_{\mu\nu}\neq\,0\}$
reads
\begin{equation}
\label{Gsol}
e^{-\lambda}
=
\left(1-\frac{2\,M}{r}\right)
\left[1+\alpha
\left(\frac{\ell}{r-B\,M}\right)^{A}\right]
\ ,
\end{equation}
which, beside the Killing and causal horizon at $\rH=2\,M$, contains a possible real singularity,
as we briefly discuss below. In fact, the physical content of the system is clarified by the explicit computation
of the effective density, 
\begin{equation}
\tilde{\rho}
=
\theta_0^{\ 0}
=
-\frac{\alpha}{k^2\,r^2}
\left(\frac{\ell}{r-B\,M}\right)^A
\left[1-A\left(\frac{r-2\,M}{r-B\,M}\right)\right]\ ,
\label{Gefecden}
\end{equation}
the effective radial pressure
\begin{equation}
\tilde{p}_{r}
=
-\theta_1^{\ 1}
=
\frac{\alpha}{k^2\,r^2}
\left(\frac{\ell}{r-B\,M}\right)^A
\ ,
\label{Gefecprera}
\end{equation}
and the effective tangential pressure
\begin{equation}
\tilde{p}_{t}
=
-\theta_2^{\ 2}
=
-\frac{\alpha\,A}{2\,k^2\,r^2\,\ell}
\left(\frac{\ell}{r-B\,M}\right)^{A+1}
\left(r-M\right)
\ .
\label{Gefecptan}
\end{equation}
We see that the effective density and effective pressures diverge at 
\begin{equation}
\label{diver}
r_{\rm c}
=
B\,M
\ ,
\end{equation}
which represents a true singularity at $0<r_{\rm c}<\rH$ (therefore hidden inside the horizon)
for $0<B<2$, that is
for
\begin{equation}
b<0
\quad
{\rm or}
\quad
b>4
\ .
\end{equation}
On the other hand, for $B>2$ (equivalently, for $0<b<2$), this singularity occurs outside the Killing horizon,
$r_{\rm c}>\rH$, and the system would represent a naked singularity.
However, this case could still represent the exterior space-time for a compact source of radius $R>r_{\rm c}$. 
\section{Einstein-Klein-Gordon system}
\label{s4}
In the previous Section we considered a generic energy-momentum tensor $\theta_{\mu\nu}$
satisfying the linear equation of state~\eqref{generic}, for which the tensor-vacuum
$\{T_{\mu\nu}=0,\,\theta_{\mu\nu}\neq\,0\}$ generates the MGD metric~\eqref{Gsol}.
We now wish to consider the specific case of a static scalar field $\psi=\psi(r)$ minimally coupled
with gravity and with a self-interaction potential $V=V(\psi)$.
\par
Let us recall the action for this scalar field $\psi$ in a curve space-time is given by
\begin{eqnarray}
S
\!\!&\!\!=\!\!&\!\!
\int\left[\frac{1}{2}\,\nabla_\mu\psi\nabla^{\mu}\psi-V(\psi)\right]
\sqrt{-g}\,d^{4}\,x
\nonumber
\\
\!\!&\!\!\equiv\!\!&\!\!
\int
{\mathcal L}\,
\sqrt{-g}\,d^{4}\,x
\ ,
\label{KGaction}
\end{eqnarray}
whose Euler-Lagrange equation for the scalar field, 
\begin{equation}
\label{KGeuler}
\frac{\partial{\cal L}}{\partial\psi}
-
\frac{1}{\sqrt{-g}}\,\partial_\mu
\left(\sqrt{-g}\,\frac{\partial{\cal L}}{\partial\,\partial_\mu\psi}\right)
=
0
\ ,
\end{equation}
is just the KG equation
\begin{equation}
\label{KGeqmotion}
\nabla_\mu\nabla^{\mu}\psi+\frac{dV}{d\psi}
=
0
\ .
\end{equation}
\par
The energy-momentum tensor associated with the scalar field $\psi$ is given by
\begin{eqnarray}
\theta_{\mu\nu}
\!\!&\!\!=\!\!&\!\!
2\,\frac{\partial{\cal L}}{\partial g^{\mu\nu}}-g_{\mu\nu}\,{\cal L}
\nonumber
\\
\label{KGem2}
\!\!&\!\!=\!\!&\!\!
\nabla_\mu\psi\,\nabla_\nu\psi
-\left(\frac{1}{2}\,\nabla_\alpha\psi\,\nabla^\alpha\psi-V\right)
g_{\mu\nu}
\ .
\end{eqnarray}
Upon inserting it in the Einstein's field equations~\eqref{ec1}-\eqref{ec3}, we obtain the general
Einstein-KG system
\begin{eqnarray}
\label{ec1kg}
k^2
\left[ T_0^{\ 0}
+\frac{1}{2}\,e^{-\lambda}\,{\psi'}^2+V\right]
\!\!&\!\!=\!\!&\!\!
\strut\displaystyle\frac 1{r^2}
-e^{-\lambda }\left( \frac1{r^2}-\frac{\lambda'}r\right)
\\
\label{ec2kg}
k^2
\strut\displaystyle
\left[T_1^{\ 1}-\frac{1}{2}\,e^{-\lambda}\,{\psi'}^2+V\right]
\!\!&\!\!=\!\!&\!\!
\frac 1{r^2}-e^{-\lambda }\left( \frac 1{r^2}+\frac{\nu'}r\right)
\\
\label{ec3kg}
k^2
\strut\displaystyle
\left[T_2^{\ 2}+\frac{1}{2}\,e^{-\lambda}\,{\psi'}^2+V\right]
\!\!&\!\!=\!\!&\!\!
-\frac {e^{-\lambda }}{4}
\left(2\nu''+\nu'^2
\phantom{\frac{A}{A}}
\right.
\nonumber
\\
&&
\left.-\lambda'\nu'
+2\,\frac{\nu'-\lambda'}r\right)
\ ,
\end{eqnarray}
along with the conservation equation~\eqref{con11}, which reads 
\begin{eqnarray}
\label{conKG}
&
\left({T}_1^{\ 1}\right)'
-
\frac{\nu'}{2}\left({T}_0^{\ 0}-{T}_1^{\ 1}\right)
-
\frac{2}{r}\left({T}_2^{\ 2}-{T}_1^{\ 1}\right)
&
\nonumber
\\
&
+e^{-\lambda}\psi'
\left[\psi''-\frac{\lambda'}{2}\psi'
-\frac{d V}{d\psi}+\left(\frac{\nu'}{2}+\frac{2}{r}\right)\psi'\right]
=
0
\ .
&
\end{eqnarray}
For the particular case of the ``scalar-vacuum" $\{T_{\mu\nu}=0,\,\theta_{\mu\nu}\neq\,0\}$,
the MGD-decoupling yields the Schwarzschild deformed metric~\eqref{Schw}, whose deformation
$f^{*}$ satisfies the set of Eqs.~\eqref{ec1d}-\eqref{ec3d}, which in this case reduces to
\begin{eqnarray}
\label{kg1d}
k^2
\left[\frac{1}{2}\,e^{-\lambda}\,{\psi'}^2+V\right]
\!\!&\!\!=\!\!&\!\!
-\frac{\alpha\,f^{*}}{r^2} 
-\frac{\alpha\,f^{*'}}{r}\ ,
\\
\label{kg2d}
k^2
\left[-\frac{1}{2}\,e^{-\lambda}\,{\psi'}^2+V\right]
\!\!&\!\!=\!\!&\!\!
-\alpha\,f^{*}\left(\frac{1}{r^2}+\frac{\nu'}{r}\right)\ ,
\\
\label{kg3d}
k^2
\left[\frac{1}{2}\,e^{-\lambda}\,{\psi'}^2+V\right]
\!\!&\!\!=\!\!&\!\!
-\frac{\alpha\,f^{*}}{4}\left(2\,\nu''+\nu'^2+2\frac{\nu'}{r}\right)
\nonumber
\\
&&
-\frac{\alpha\,f^{*'}}{4}\left(\nu'+\frac{2}{r}\right)
\ ,
\end{eqnarray}
and the conservation equation~\eqref{conKG} reads
\begin{equation}
\label{conKG2}
\psi''+\left[\frac{2}{r}+\frac{1}{2}\left(\nu'-\lambda'\right)\right]\psi'
=
e^{\lambda}\,\frac{dV}{d\psi}
\ ,
\end{equation}
which is just the explicit form of the KG equation~\eqref{KGeqmotion} for the
static and spherically symmetric metric~\eqref{metric}.
\par
Let us recall that the temporal metric component in~\eqref{pfmetric} is not deformed under the MGD.
Hence the system~\eqref{kg1d}-\eqref{conKG2} contains the three unknown functions $\{f^{*},\,\psi,\,V\}$
to be determined by three independent equations among Eqs.~\eqref{kg1d}-\eqref{conKG2}.
No further restriction on the system is therefore necessary. 
\par
Despite the above, we can obtain some useful information from the generic equation of state~\eqref{generic}
introduced to close the system~\eqref{ec1d}-\eqref{ec2d}.
By simple inspection of Eqs.~\eqref{kg1d}-\eqref{kg3d}, we notice that the energy-momentum tensor
$\theta_{\mu\nu}$ associated with the scalar field $\psi$ satisfies an equation of state like that in
Eq.~\eqref{generic} with $\{a=0,\,b=1\}$, which yields $\{A=2,\,B=3\}$
in the expressions~\eqref{A} and~\eqref{B}~\footnote{We notice though that $a=0$ makes such an equation
of state singular, since we cannot express $\theta_1^{\ 1}\sim \tilde p_r$ as a function of $\tilde\rho$.
This is the reason we cannot guarantee all scalar field solutions can be expressed in the form given
in this paper.}. 
Therefore, according to the expression~\eqref{giso2}, the scalar field $\psi$ will produce a geometric
deformation on the Schwarzschild vacuum given by
\begin{equation}
f^*(r)
=
\left(1-\frac{2\,M}{r}\right)
\left(\frac{\ell}{r-3\,M}\right)^{2}
\ ,
\label{KGdef}
\end{equation}
where $\ell$ is a constant with dimensions of a length.
Also, as stated in Eq.~\eqref{diver}, a naked singularity should appear at $r_{\rm c}=3\,M$.
We can see this singularity for the values $\{A=2,\,B=3\}$ in the generic solution for the tensor-vacuum
in Eq.~\eqref{Gsol}, namely
\begin{equation}
\label{GsolKG}
e^{-\lambda}
=
\left(1-\frac{2\,M}{r}\right)
\left[1+\alpha
\left(\frac{\ell}{r-3\,M}\right)^{2}\right]
\ .
\end{equation}
Indeed, the Ricci scalar $R_{\ \alpha}^{\alpha}$ and the Ricci-Ricci scalar $R_{\alpha\beta}R^{\alpha\beta}$
present a singularity~\footnote{The expression for the Kretschmann scalar
$R_{\alpha\beta\gamma\delta}R^{\alpha\beta\gamma\delta}$
is too large to display here, but it also diverges for $r=r_{\rm c}=3\,M$.}
at $r=r_{\rm c}=3\,M$, 
\begin{eqnarray}
R_\alpha^{\,\,\alpha}
\!\!&\!\!=\!\!&\!\!
-\frac{2\,\alpha\,\ell^2}{(r-3\,M)^3\,r}
\\
R_{\alpha\beta}R^{\alpha\beta}
\!\!&\!\!=\!\!&\!\!
\frac{4\,\alpha^2\,\ell^4(3\,M^2-3\,M\,r+r^2)}{(r-3\,M)^6\,r^4}\ .
\end{eqnarray}
\par
We must conclude that the solution~\eqref{GsolKG} cannot represent a black hole but the exterior geometry
of a self-gravitating system of radius $R>r_{\rm c}$, surrounded by a spherically symmetric minimally coupled
scalar field $\psi$, which satisfies 
\begin{equation}
\label{dpsi}
{\psi'}^2
=
\frac{2\,\alpha\,\ell^2}{k^2\,(r-3\,M)\,r\,\left[\alpha\,\ell^2+(r-3\,M)^2\right]}
\ ,
\end{equation}
and whose potential $V$ is given by
\begin{equation} 
\label{VKG}
V=\frac{\alpha\,\ell^2\,M}{k^2\,r^2\,(r-3\,M)^3}\ .
\end{equation}
\par
In order to find the functional $V=V(\psi)$, we consider the expressions in Eq.~\eqref{dpsi} at first order in $\alpha$,
hence we have  
\begin{equation}
\label{psi}
{\psi'}^2
\simeq
\frac{2\,\alpha\,\ell^2}{k^2\,r\,(r-3\,M)^3}
\ ,
\end{equation}
which yields (for simplicity, we set the integration constant to zero)
\begin{equation}
\label{psi2}
{\psi}^2
\simeq
\frac{8\,\alpha\,\ell^2}{9\,k^2\,M^2}\frac{r}{(r-3\,M)}
\ .
\end{equation}
Using the expression~\eqref{psi2} in Eq.~\eqref{VKG} we obtain~\footnote{Note that this expression is regular
for $\alpha\to 0$ when computed for the solution $\psi^2\sim \alpha$, but the limit becomes singular off-shell.}
\be
V
\simeq
\frac{\alpha\,\ell^2}{243\,k^2\,M^4}\,\frac{\psi^6}{K^6}\left(1-\frac{K^2}{\psi^2}\right)^5
\,;
\ 
K^2\equiv\,\frac{8\,\alpha\,\ell^2}{9\,k^2\,M^2}
\ .
\label{VKG2}
\ee
\par
The scalar field $\psi$ fills the Schwarzschild vacuum with an effective density 
\begin{equation}
\tilde{\rho}
=
\theta_0^{\ 0}
=
\frac{\alpha}{k^2\,r^2}
\frac{\ell^2}{\left(r-3\,M\right)^3}(r-M)\ ,
\label{KGefecden}
\end{equation}
an effective radial pressure
\begin{equation}
\tilde{p}_{r}
=
-\theta_1^{\ 1}
=
\frac{\alpha}{k^2\,r^2}
\left(\frac{\ell}{r-3\,M}\right)^2
\ ,
\label{KGefecprera}
\end{equation}
and an effective tangential pressure
\begin{equation}
\tilde{p}_{t}
=
-\theta_2^{\ 2}
=
-\tilde{\rho}\ .
\label{KGefecptan}
\end{equation}
\par
\begin{figure}[t]
\center
\includegraphics[scale=0.3]{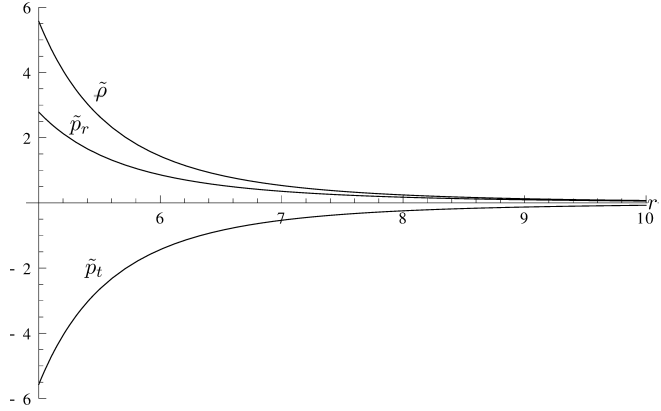}
\\
\centering\caption{Effective variables $\{\tilde{\rho},\,\tilde{p}_r,\,\tilde{p}_t\}\times 10^4$
around a self-gravitating system of radius $R=5$ and mass $M=1$ for $\alpha\,\ell^2=0.7$.}  
\label{f1}      
\end{figure}

We see that the dominant energy condition for both pressures, namely, $\tilde{\rho}\geq\,|\tilde{p}_r|$
and $\tilde{\rho}\geq\,|\tilde{p}_t|$, is satisfied. Figure~\ref{f1} shows the density and pressures
in~\eqref{KGefecden}-\eqref{KGefecptan} respectively for $\alpha\,\ell^2=0.7$.
\section{Solar System classical tests}
\label{s5}
Results from solar system tests remain among the pivotal observational evidences of
GR.
For this reason, previous developments obtained with the MGD approach were confronted with the 
classical tests in Ref.~\cite{jo12}, thereby deriving upper bounds for the MGD parameters.
More recently, relevant observational results from the lensing in the strong field limit were
included~\cite{roldaoGL}. 
Now, we first observe that the solution in Eq.~(\ref{GsolKG}) in fact contains just one free parameter
$\mathfrak{L}^2\equiv \alpha\,\ell^2$, which we will place bounds for, by analysing the perihelion precession
of Mercury together with the light deflection by the Sun.
\subsection{Perihelion Precession}
Orbits in a static spherically symmetric space-time are usually described by employing $u=1/r$
so that the geodesic equation reads~\cite{boemer}
\begin{eqnarray}
\frac{d^{2}u}{d\phi ^{2}}\!+\!u=k(u)
\ .
\end{eqnarray} 
For the metric~\eqref{Schw} with the MGD in Eq.~\eqref{GsolKG}, one finds 
\begin{eqnarray}
k(u)
\!\!&\!\!=\!\!&\!\!
\frac{1}{2\,L^2\,(1-2\, M\, u)^2}
\nonumber
\\
&&
\times
\left\{6\,L^2\, M\, u^2\, (1-2\, M\, u)^2
-\mathfrak{L}^2 \left(\frac{1}{u}-3 \,M\right)^{1/2}\right.
\nonumber
\\
&&
\left.
\times
\left[-\frac{E^2}{2}+L^2\, u^2 \left(2\,M\, u\left\{9\, M\, u-\frac{7}{2}\right\}\right.\right.\right.
\nonumber
\\
&&   
\left.\left.\left.
\phantom{\frac{A}{B}}
-6\, M\,u-5\right)\right]\right\}
\ ,
\label{kudef}
\end{eqnarray}
where $E$ is the energy, $L={2\,a^{2}\pi}\,\sqrt{1-e^{2}}/{T}$ the angular momentum,
the period $T$ and eccentricity $e$.
\par
One denotes by $\gamma^2(u)=1-k'$, where $k'={dk}/{du}$, a scalar to be evaluated on the 
circular orbit determined by $u=u_{0}$, $u_0$ being a root of $u_{0}=k(u_{0})$.
For small values of $k'(u_{0})$, the perihelion advances by $k'(u_{0})/2$~\cite{jo12}.
Any further deviation out of the circular orbit is ruled by
$\delta/\delta_0 \sim \sin\left(\gamma(u)\phi +\pi/2+\uptheta \right)$, where $\delta _{0}$
and $\uptheta$ are constants~\cite{boemer}.
In the geometry described by the metric element~\eqref{GsolKG}, a complete orbit corresponds
to an angle $\phi ={2\,\pi }/{\gamma(u)}$, with 
\begin{eqnarray}
&&
\gamma(u)
=  
\frac{M \left(\frac{1}{u}-3\, M\right)^{-1/2}}{4\,L^2\, u\, (2\, M\, u-1)}
\label{gamma}
\\
&&
\times
\left\{L^2\, u^2\left[2\,\mathfrak{L^2}\,(2-3\, M\, u)+2\,u
\left(6\,\{2\, M\,u-1\}
\right.\right.\right.
\nonumber
\\
&&
\left.\left.\left.
\times\left\{\left[\frac{1}{u}-2\,M\right]^{\frac{1}{2}}
+\mathfrak{L^2}\right\}
-\mathfrak{L^2}\right)
\right]
-E^2\,\mathfrak{L^2}
\right\}
\ .
\nonumber
\end{eqnarray}
The perihelion advance reads $\delta \phi=\delta \phi _{\rm GR}-{f}(\mathfrak{L},a_0,M)$,
where $\delta \phi _{\rm GR}=6\,\pi\,M/a\left(1-e^{2}\right)$ denotes the Schwarzschild
precession formula and ${f}(\mathfrak{L},a_0,M)$ is a correction that can be estimated
numerically from the above Eq.~\eqref{gamma}.
By converting to the usual SI units, in which the speed of light $c=2.99792\times 10^{8}~{\rm m/s}$,
the Newton constant $G_{\rm N}=6.67429\times 10^{-11}~{\rm m^{3}/Kg \,s^{2}}$,
the Sun mass $M_{\odot}=1.98855\times 10^{30}~{\rm kg}$ and the Sun radius
$R_{\odot}=6.95660 \times 10^8~ {\rm m}$,
$a=5.79112\times 10^{10}~{\rm m}$ and $e=~2.056154\times 10^{-1}$, 
it yields $\delta{\phi}-\delta \phi _{\rm GR}=0.13\pm 0.22\,$arcsec/century.
Upon comparing with experimental data, this yields a bound
\begin{equation}
\mathfrak{L}
\lesssim
(1.21 \pm1.34)\times 10^{-7}\,{\rm m}.
\label{Lprec}
\end{equation}%
\subsection{Light Deflection}
Photons travel along null geodesics, whose equation of motion reads 
\begin{equation}
\frac{d^{2}u}{d\phi ^{2}}+u
=
\frac{1}{2}\frac{dp(u)}{du}
\ ,
\end{equation}
where $p(u)=e^{-\nu -\lambda}\,(E^2/L^2)+g(u)\,u^{2}$,  and $g(u)=1-e^{-\lambda}$.  
The lowest order approximation yields $u={\cos \phi }/{a_0},$ where $a_0$ is the distance of
closest approach to the stellar distribution of mass $M$. 
The total deflection angle of light rays, $\delta =2\,\varepsilon$~\cite{boemer}, for the
geometry in Eq.~(\ref{GsolKG}) yields $g(u)=2\,M\,u$, resulting in
\begin{eqnarray}
\!\!\!\!
&&
p(u)
=
\frac{a_0-3\,M}{2\,L^2\,(1-2\,M\, u)^2}
\left\{\frac{a_0\,\mathfrak{L}^2\,u^2}{1-3\, M\,u}
\left[L^2\, u^2 \right.
\right.
\nonumber
\\
\!\!\!\!
&&
\left.
\left.\times\left(6\, M\, u-5
-E^2\,\{2\, M\, u-3\}-4\,M\,u\,\{4\,M\, u-3\}
\right)\right]
\right.
\nonumber
\\
\!\!\!\!
&&
\left.
+3\,M\,L\, u \,\frac{(1-3\, M\, u)^2}{a_0-2\, M}
\left(2\, u+1-3\,M\,u\right)
\right\}
\ .
\label{pudef}
\end{eqnarray}
The total deflection angle reads $\delta \phi \simeq {4\,M}/{a_0} + f_1(\mathfrak{L}, M, a_0)$,
where $f_1$ is a complicated function of the MGD parameter $\mathfrak{L}^2$. 
The best available data regarding light deflection by the Sun come from long baseline radio interferometry,
yielding $\delta \phi = \delta \phi_{\rm GR} (1 + \kappa)$, where $\kappa\lesssim 1\times 10^{-3}$, for  
$\delta \phi_{\rm GR} \sim 1.75$ arcsec.
In the limit $\left({M}/{a_0}\right)^2\ll 1$, ${E^2}-1\ll1$ and ${M}\ll L$, it numerically implies that
\begin{equation}
\label{Llight}
\mathfrak{L}
\lesssim
(1.34\pm5.36 ) \times 10^{-7}\, {\rm m},
\end{equation}
which is (slightly) less constraining than Eq.~\eqref{Lprec} due to the larger error.
\section{Conclusions}
\label{s6}
\setcounter{equation}{0}
By making use of the MGD-decoupling approach, we have presented in detail how the Schwarzschild metric
is modified when the vacuum is filled by a spherically symmetric KG scalar field $\psi$,
which gravitates but does not interact directly with the matter that sources the Schwarzschild solution.
\par
We found that the scalar field deforms the Schwarzschild solution in such a way that 
a free parameter $\mathfrak{L}^2\equiv \alpha\,\ell^2$, which encodes the anisotropy associated
with $\psi$, appears explicitly in the new solution, as we see in Eq.~\eqref{GsolKG}.
The main characteristic of this solution is that it shows a singularity at $r=r_{\rm c}>2\,M$, and it cannot
therefore represent a black hole but the exterior geometry of a self-gravitating system of radius $R>r_{\rm c}$,
surrounded by a spherically symmetric minimally coupled scalar field $\psi$.
This scalar field defines a fluid whose density and pressures satisfy the dominant energy conditions,
and whose main consequence is that it weakens the gravitational field of the central source of the
Schwarzschild metric.
\par
By using two Solar System classical tests, namely, the perihelion precession and the light deflection,
we bounded the value of the free parameter $\mathfrak{L}$ associated with the scalar-vacuum.
These values, presented in Eqs.~\eqref{Lprec} and~\eqref{Llight}, show that the perihelion precession
gives a more stringent upper bound for this parameter, namely,
$\mathfrak{L}\lesssim\,(1.21 \pm1.34)\times 10^{-7}\,{\rm m}.$
\par
Finally, we want to notice that since the temporal metric component in Eq.~\eqref{metric} is not deformed
under the MGD, the field equations for the tensor-vacuum $\{T_{\mu\nu}=0,\,\theta_{\mu\nu}\neq\,0\}$,
namely the ``quasi-Einstein'' system in Eqs.~\eqref{ec1d}-\eqref{conset2}, contains four unknown
functions $\{{\theta}_0^{\ 0},\,{\theta}_1^{\ 1},\,{\theta}_2^{\ 2},\,f^{*}\}$ to be determined by three independent
equations in the system~\eqref{ec1d}-\eqref{conset2}.
Therefore we need to introduce an additional restriction on the tensor-vacuum to solve it, as indeed
was developed in Ref.~\cite{MGDBH2}.
However, in the case of the scalar-vacuum~\eqref{kg1d}-\eqref{conKG2}, we have three unknown
functions $\{f^{*},\,\psi,\,V\}$ to be determined by three independent equations among those in
Eqs.~\eqref{kg1d}-\eqref{conKG2}.
Hence, no further restriction is necessary.
The above is very significant, since it shows that under the MGD-decoupling, the Schwarzschild vacuum
has associated a specific scalar configuration $\{\psi,\,V\}$, which will produce a specific anisotropy through
the deformation $f^{*}=f^{*}(r)$ in Eq.~\eqref{KGdef}. 
We conclude that there is no black hole solutions in the Einstein-KG system under MGD-decoupling.
However, the extension of the MGD-decoupling based in the extension of the MGD-approach, as developed
in Ref.~\cite{MGDextended1}, could still yield black hole solutions.
This development goes beyond the objective of this article.
\section*{acknowledgments}
\par
J.O.~and S.Z.~have been supported by the Albert Einstein Centre for Gravitation and Astrophysics financed
by the Czech Science Agency Grant No.14-37086G.
R.C.~is partially supported by the INFN grant FLAG and his work has been carried out in the framework
of GNFM and INdAM and the COST action {\em Cantata\/}.
R.dR.~is grateful to CNPq (Grant No. 303293/2015-2), and to FAPESP (Grant No. 2017/18897-8)
for partial financial support.
A.S.~is supported by Project Fondecyt 1161192, Chile.
%

%

%
%
%%%%%%%%%%%%%%%%%%%%%%%%%%%%%%%%%%%%%%%%%%%%%%%%%%%%%%%%%%%%%%%%%%%
%%%%%%%%%%%%%%%%%%%%%%%%%%%%%%%%%%%%%%%%%%%%%%%%%%%%%%%%%%%%%%%%%%%
\end{document}